# Scalable High Performance SDN Switch Architecture on FPGA for Core Networks

Sasindu Wijeratne, Ashen Ekanayake, Sandaruwan Jayaweera, Danuka Ravishan and Ajith Pasqual
Dept. of Electronic and Telecommunication Engineering, University of Moratuwa, Sri Lanka
{130665u, 130150l, 130254j, 130516p}@uom.lk, pasqual@mrt.ac.lk

*Abstract*—Due to the increasing heterogeneity in network user requirements, dynamically varying day to day network traffic patterns and delay in network service deployment, there is a huge demand for scalability and flexibility in modern networking infrastructure, which in return has paved way for the introduction of Software Defined Networking (SDN) in core networks. In this paper, we present an FPGA-based switch which is fully compliant with OpenFlow; the pioneering protocol for southbound interface of SDN. The switch architecture is completely implemented on hardware. The design consists of an OpenFlow Southbound agent which can process OpenFlow packets at a rate of 10Gbps. The proposed architecture speed scales up to 400Gbps while it consumes only 60% resources on a Xilinx Virtex-7 featuring XC7VX485T FPGA.

## I. INTRODUCTION

Software Defined Networking (SDN) is a novel networking model which facilitates innovation in the field of networking. SDN is more dynamic, adaptable and manageable, which makes it ideal for the high-bandwidth and dynamic nature of today's applications.

Recent studies have shown that SDN has the potential to be used in a vast variety of network types [1]. In decoupled SDN architecture, infrastructure layer of SDN is made of switches which are dedicated to forwarding the traffic according to the flow details stored in their flow tables. This flow tables are maintained by the control layer. OpenFlow is the most common Southbound API which is used as the medium to communicate between control layer and infrastructure layer. It defines all possible functions and messages used to manage switches via a centralized software controller.

With uprising cutting edge technologies for the core net-work, SDN has shown promising results in efficient routing techniques for service provider networks and advancements of ISP backbone. Also, SDN approach promises reduction of costs, enhancement of network flexibility and scalability, and shortening of the time to market of new applications and services which is much needed for emerging core network requirements.

Recent researches on SDN switch implementations are focused on Field Programmable Gate Arrays (FPGAs) considering the scalability, flexibility and low cost [2][3][4][5]. Khan et al [5] has illustrated a hardware level SDN switch which meets the 10Gbps line rate and it is fully-compliant with OpenFlow 1.0. This system has successfully off-loaded to NetFPGA-10G, ML605 and DE04 FPGA boards. Zhou et al. in [2] has demonstrated a flexible and scalable hybrid switch (software and hardware level switch implementation on a SoC). This scalable architecture has achieved 88Gbps with a 1000 flow table design. For lookup process, this implementation has used Ternary Content Addressable Memory (TCAM) and a single table structure. The OpenFlow Agent of this design is purely software and it is capable of decoding OpenFlow 1.0. Wijekoon et al. [3] illustrated a hardware level SDN switch with scalability, which is focused on low cost and efficient flow matching. This system is implemented on Virtrex-7 FPGA. Naous et al. in [4] describes an OpenFlow switch on NetFPGA platform. It has the capability of handling 32,000 exact match flow entries and running at line-rate with four 1G NetFPGA ports.

In this paper, we present a novel high-performance switch fabric with low-latency for SDN environment which is based on OpenFlow [6]. This cost-efficient design is focused on re-configurable hardware and it can be scaled with the network's requirements. Further, the proposed architecture provides high throughput and high-speed connection with control plane which makes it an ideal candidate for core networks.

Section II briefly describes the background of our research including the information regarding data plane of SDN. In section III, we explain the design space exploration for our architecture. Finally, in section IV, a performance evaluation of the switch fabric along with results obtained for this architecture are provided.

## II. BACKGROUND

The data plane comprises switches with flow tables which contain flow entries that include routing paths to the incoming network packets. A flow entry consists of packet header fields, counters, a priority and associated actions. The SDN control layer is responsible for flow entry maintenance. The bridge between control layer and a switch is the Southbound interface. OpenFlow is the southbound interface protocol in this implementation. The rich feature list of OpenFlow makes it more suitable for the dynamic environment. In switch, OpenFlow Southbound agent is responsible for the communication with control layer. The implemented OpenFlow agent is capable of handling messages from OpenFlow 1.3.1 and above [6].

Exact matching, range matching and prefix matching mechanisms are used in flow tables. The efficiency of matching process directly affects the efficiency and speed of traffic forwarding. Ternary Content Addressable Memory (TCAM) and Content Addressable Memory (CAM) make the matching process more efficient even though the hardware utilization is high.

Each flow entry has its own priority given by the controller and those who have higher priority values get examined first.

Duration of the flow entry, packet count and byte count received are maintained by the counters. Further, per table, per port and per queue counters are there to record the condition of the switch. These counter values help control layer to make better and efficient forwarding decisions.

Implementing a resource efficient solution which can support all these functionalities while handling the core network throughput and latency requirements is the focus of our research.

## III. ARCHITECTURE

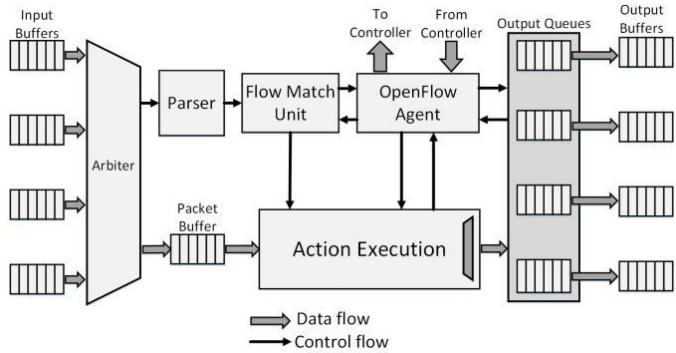

Fig. 1. The Overall Architecture of the SDN switch IP

The figure 1 shows the overall architecture of the proposed SDN switch IP. This design is a full hardware/FPGA implementation which connects with the SDN Controller through a secure channel using OpenFlow Agent. This agent acts as the main OpenFlow translator component of the southbound interface inside the switch design.

As shown in figure 1, when a packet arrives to the switch, it is queued in an input buffer. Then an Arbiter chooses a packet from the input buffers while considering the traffic of each input port. Arbiter passes the chosen packet to the internal Packet Buffer while forwarding a copy of packet header to the Parser in order to extract the header fields. Parser creates a tuple with packet forwarding information which is extracted from corresponding header fields. This tuple is forwarded to the Flow Match Unit where it is matched against existing flow entries using exact match method or wild card method. Flow entries are maintained under the guidance of the OpenFlow Controller. Then the match information is passed to Action Execution Unit. It takes the corresponding packet from internal Packet Buffers and forward it to the corresponding queue of the output ports.

### A. Packet Parser

Packet Parser is responsible for packet header field extraction and passing the extracted information to Flow Match Unit in order to take decisions about corresponding packet. The implemented parser engine is operated on a replicated version of the network headers of each packet. In accordance with the parser graph supported by the switch design, the packet-header possesses the information for the end to end routing of a packet through a network. This packet-header size can be reconfigured to support different parser graph implementations if required. Unlike in a pipelined Parser design where separate pipeline stages are implemented to extract the data of different network headers in separate clock cycles, complex combinational logic is used by the proposed Parser architecture to extract separate headers in a single clock cycle. Still, the extractions are done by separate modules which can be easily reconfigured to extract different length headers. These extraction modules can be categorized into two types according to the field types (dynamic or constant length fields) contained in the headers extracted. At the end of each parser extraction, a selected set of extracted fields are sent to the Flow Match Unit to match with the existing flow entries.

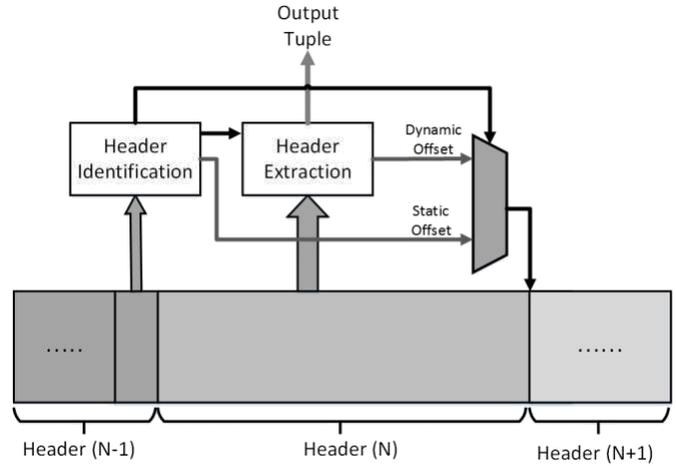

Fig. 2. Dynamic Header Extraction

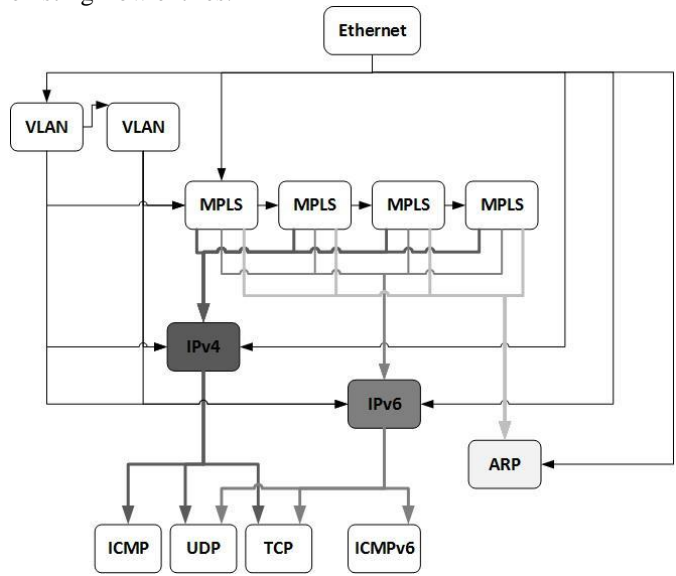

Fig. 3. Parser graph

Fields such as Ether Type and IPV4 Protocol are essential

for the Parser engine in its field extraction process to direct the engine to the next parsing logic. Therefore, the parser design is generated with the ability to extract and identify any value these fields have to offer. The design can decode most combinations of the core network can offer. If the Parser receives any contradicting data from a packet, it identifies the packet as malicious and sends a signal to the Flow Match Unit to drop it.

Unlike a fixed length header, a header like an IPv4 header of a packet can have a dynamic length. The length of the header is sent along with the IPv4 header. The header length resides in the Internet Header Length (IHL) field. As a result of the dynamic length of the IPv4 header, the placement of the next header can vary. Therefore, the extraction of this next header is done using shift registers. They dynamically shift the bits for a specified length to capture the data. Additionally, the VLAN tags and MPLS tags are also extracted by the Parser if they appear in packets.

*B. Flow Match Unit*

The Flow Match Unit maintains the flow tables and its entries while maintaining the flow statistics of each data flow and flow table. Main functionalities of this unit are match the input flows with the existing flow entries inside the tables, write flow entries to flow tables, assign actions and instructions to each flow entry and Maintain statistic counters.

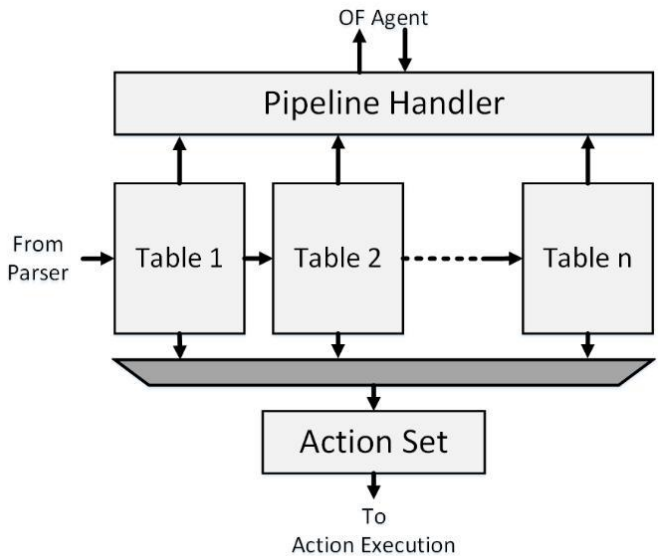

Fig. 4. Overview of Match Action Unit

This hardware intensive unit consists of multiple flow tables. A tuple containing header fields and input port data of a packet is passed to the Flow Match Unit from the Parser. This unit is responsible for finding corresponding action set to each input flow coming through the Parser. Flow matching mechanism always starts with the first flow table and it propagates through the table pipeline by "goto <table id>" instruction. When jumping to the next table, table id should always be higher than the previous which indicates no backward propagation through the pipe. Propagation stops when it hits the flow entry which does not has a "goto <table id>" instruction. Then the accumulated action set is passed to the Action Execution. Figure 4 shows an overview of the hardware implementation of the Flow Match Unit architecture.

Wild-card lookup is required for some of the match fields such as IPv4 addresses. TCAM modules are implemented inside the flow table modules to support wild-card matching while CAM modules are implemented to support exact matching. CAMs and TCAMs can match the input tuple with the existing flow entries within one clock cycle where adding a new flow entry into them takes several clock cycles. Both TCAM and CAM use dual port RAMs (DP-RAM) to store the flow entries because the write and read operations can be done via both interfaces of DP-RAM at the same time in different addresses.

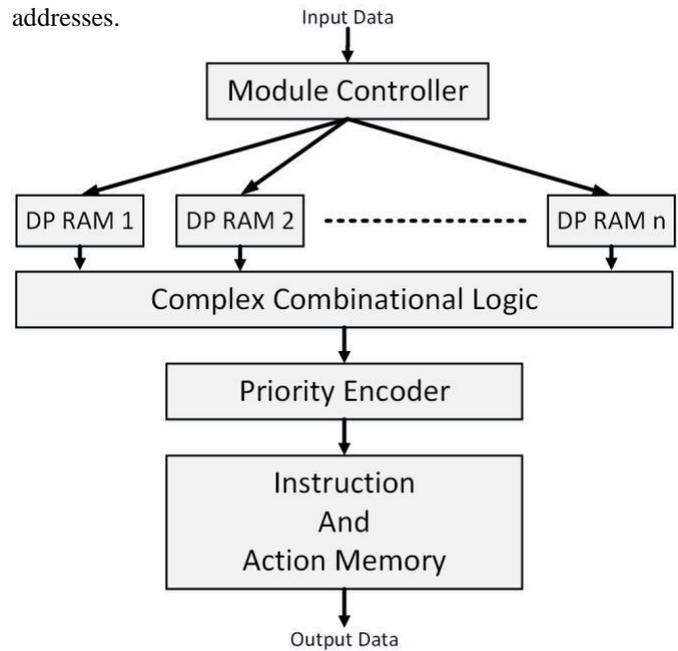

Fig. 5. CAM Architecture

As shown in the figure 5, the CAM architecture is using multiple DP-RAMs to store the flow entries. They are stored as addresses of the RAMs [7] [8]. First, the tuple from the Parser is forwarded to the Module Controller of the Flow Match Unit. Apart from that, Module Controller is responsible for writing new flow entries to the CAM according to the Controller instructions which passes through OpenFlow South-bound Agent. As shown in figure 5, after the matching process, the output of the Priority Encoder which is the selected flow to the given tuple is mapped to the address of the Instruction and Action Memory. TCAM architecture is an extension of the

CAM architecture where the Complex Combinational Logic is extended to find the write address when wild card values are included in flow entries. Number of DP-RAMs is also increased in the TCAM architecture. When the OpenFlow Southbound agent queries the statistics of the flow tables from Pipeline Handler as shown in figure 4, it initiates requests to collect data from flow tables. A flow table maintains byte and packet counters per flow as well as per table and it replies whenever the Pipeline Handler requests for statistics.

*C. Action Execution Engine*

Action Execution Engine is the module where the action set discovered from the flow table pipeline is executed. Figure 6 shows a top-level abstract view of the architecture. In an action set, four main actions can be identified [6].

- Forward: forward the packet to any physical or virtual port
- Buffer: buffer the packet if it does not match and send the details to the controller
- Modify: modify the packet header
- Drop: drop the packet

Forward action forwards the packet which is buffered either in original packet buffer or in internal buffers to the corresponding queue of the output port. When the packet has a mismatch in the table pipeline, Controller may specify an action to send the packet details to the Controller. It comes as a form of forwarding action where the port is the Controller. Hence original packet which passes through the packet buffer should be stored inside the internal buffers until it receives corresponding actions from the Controller. In order to generate the Packet-In message [6] for table mismatch, OpenFlow Agent collects data from Flow Match Unit as well as internal packet buffers of the Action Execution engine. OpenFlow Agent receives the corresponding action set as a form of Packet-Out message [6] and passes that information to the Action Execution.

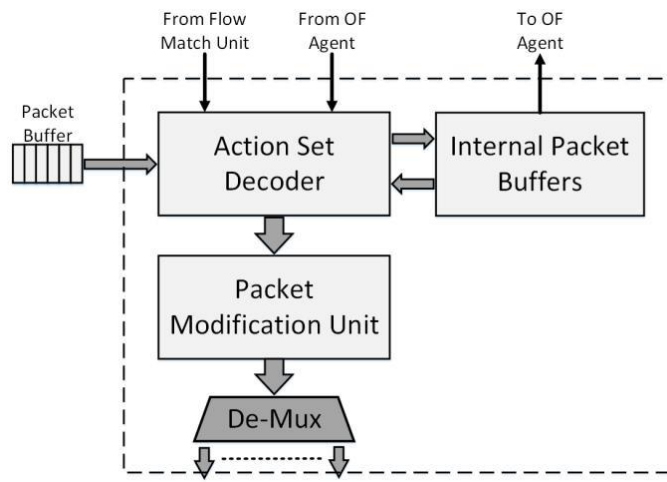

Fig. 6. Action Execution Architecture

Packet Modification Unit has the responsibility to modify the packet headers according to actions specified in the action set. Actions such as push-pop VLAN/MPLS tags, change TTL, set header field are executed here. When a packet mismatch occurs in the flow table pipeline, packet should be dropped unless the Controller has specified an action to buffer the packet. Further, if the Parser identifies a malicious packet, it informs the Flow Match Unit to drop the packet. When we deploy the switch in a new environment, flow tables are not updated and all the packets coming through the switch will have mismatches. Therefore, the buffering capacity will be exceeded since the Controller communication is comparatively slow. This will heavily impact on the packet synchronization in the switch. To overcome this, when buffer limit exceeds, it signals it to the Flow Match Unit. Then Flow Match Unit generates a packet drop action along with the action set. Hence, packet drop feature in the action execution is a vital part of packet synchronization inside the switch.

*D. OpenFlow Southbound Agent*

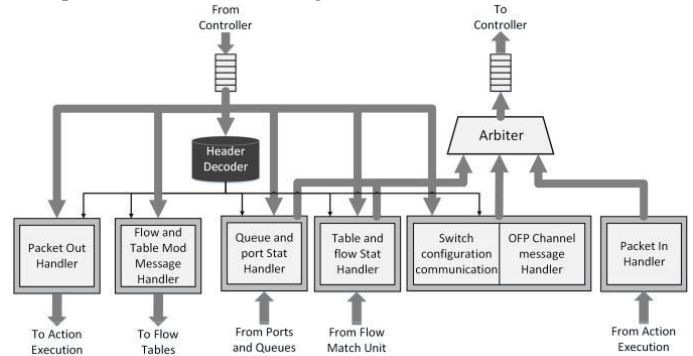

Fig. 7. The Architecture of OpenFlow Southbound Agent

The communication between Control layer and Data plane is essential for the SDN concept. For this southbound communication, we have used OpenFlow Protocol because of its regular usage in the industry and academia. The OpenFlow agent manages the communication between the Controller and the Switch from within the switch. In our proposed switch, because of the high number of dynamically changing flows in a core network environment, southbound communication is carried out using dedicated link between SDN Controller and SDN Switch. In order to withstand the high-speed traffic, OpenFlow Southbound Agent is implemented in hardware. OpenFlow agent handles five basic types of messages.

- Southbound channel setup and keep alive
- Switch information requests and replies Flow Control messages
- Packet Control messages
- Error handling messages

All the messages have the symmetry of request from Controller and reply from switch or request from switch and reply from Controller.

This module is a heavily pipelined system with low latency. OpenFlow packets from Controller are processed in-order. The data width of the connection between OpenFlow Controller and OpenFlow Agent is chosen as 8 bytes, since all OpenFlow messages are multiples of 8 bytes. According to the Open-Flow Packet (OFP) structure, first 8 bytes consists of OFP header. Initially the OFP header of each packet is decoded in Header Decoder. The OFP header contains information of the incoming packet including, OpenFlow packet type and packet length. After identifying the processing packet type and length, corresponding action module is activated, and packets are forwarded to the module from input packet buffer of OpenFlow Agent. The set of action modules of OpenFlow Agent is shown in the figure 7. The necessary actions to the packets are taken from the corresponding submodules. Those actions include:

- Extract and pass Packet Out [6] information to Action Execution Module
- Extract and pass Flow Mod and Table Mod [6] information to Flow Match Unit
- Forward current queue statistics, port statistics, table statistics and flow statistics to OpenFlow Controller

Output packets from internal modules to OpenFlow Controller are buffered module-wise. The OpenFlow Channel Output Arbiter chooses the packets in following order and passes them to the output packet buffer of OpenFlow Agent as shown in figure 7.
- Packet In [6]
- Statistics
- Switch information and configuration
- Southbound channel setup and keep alive

## IV. PERFORMANCE EVALUATION

The switch fabric is synthesized and implemented on Xilinx Virtex-7 featuring XC7VX485T FPGA using Vivado 2015.4. For the experiment, the controller plane is implemented on a host PC and connected to the FPGA through PCIe using RIFFA framework. FMC network daughter card is connected to the Xilinx FPGA evaluation kit with 4x10G ports. It is used to push traffic into the switch fabric and packet generators and packet duplicators are also implemented to simulate the high-speed network behavior.

The design works at 160 MHz internal clock frequency with TCAM and CAM having a data depth of 1K. Higher through-put requirements in core networks are achieved by increasing the internal processing data width of the architecture. With the data widths of 512, 1024, 2048 and 4096 bits we could achieve maximum throughput of 82Gbps, 164Gbps, 328Gbps and 655Gbps respectively.

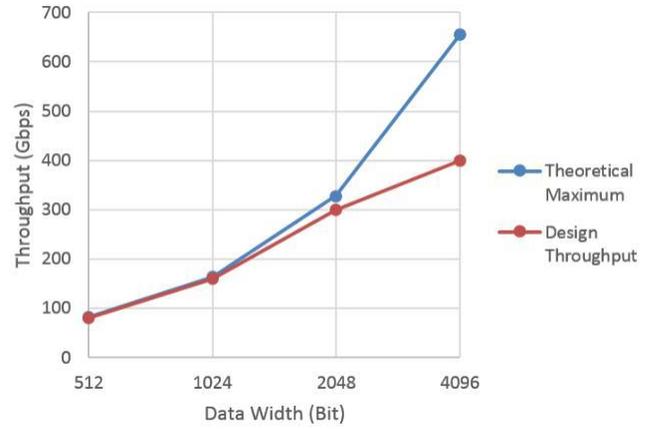

Fig. 8. Maximum throughput of the scalable switch fabric with processing data width

In figure 8 using blue, the theoretical maximum throughput of our switch fabric for each processing data width is shown. Since the switch fabric is pipelined, the maximum throughput is calculated as the product of processing data width and clock frequency. It is unable to achieve the theoretical maximum with existing property constraints of available Gigabit Ethernet MAC IPs [9]. Table I shows the properties of scaled versions of switch fabrics which are implemented on XC7VX485T-2FFG1761 FPGA. Achieved throughputs for each scaled version are shown in figure 8 using red.

TABLE I
Scalability of Switch Fabric IP

| Processing Data Width | 512 | 1024 | 2048 | 4096 |
|---|---|---|---|---|
| Clock Frequency (MHz) | 160 | 160 | 160 | 160 |
| Line Rate | 10G | 10G | 10G | 100G |
| Number of Ports | 8 | 16 | 32 | 4 |
| Throughput (Gbps) | 80 | 160 | 320 | 400 |

Figure 9 shows the resource utilization of the switch in XC7VX485T-2FFG1761 FPGA for instance in table I. The total amount of available LUTs, slice registers and BRAMs of the FPGA are 303600, 607200 and 1030 respectively.

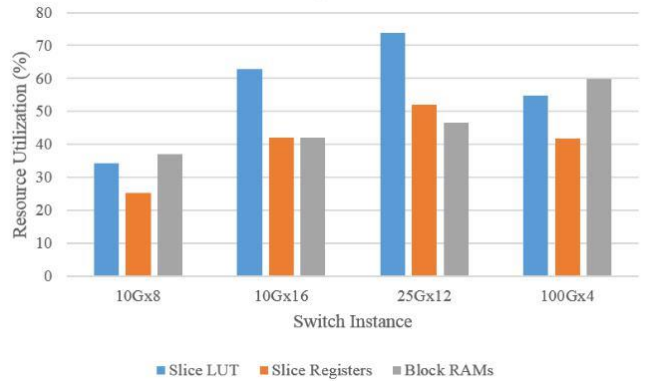

Fig. 9. Resource utilization percentage of the scalable switch fabric with processing data width

Deployment of these switch instances should be done using FPGA based embedded platforms which support enough ports and their line rates.

## V. CONCLUSION

The growth of network traffic and the varying demand for new services are severely straining the conventional networking infrastructure. With its reconfigurable design, SDN is a perfect candidate to fill this technological gap. In this paper, we have presented a flexible and scalable SDN based switch fabric on FPGA which meets the core-network requirements. The switch is a complete FPGA based architecture, implemented and experimented on Xilinx Virtex-7. OpenFlow is used as the SDN southbound protocol of the switch. Switch fabric can connect to a control plane running upon a host PC via PCIe which provides a chance at research level to explore SDN in core networks. Moreover, the architecture is experimented for different scaled versions using line rates of 10G, 25G and 100G. By using FPGA based embedded platforms which support enough ports and their line rates, this architecture can be deployed in core networks.